\documentstyle[12pt,a4wide,cite,epsf]{article}
\begin{document}
\def\ls{\vskip 12.045pt}   
\def\pn{\par\noindent}        
\def\ni{\noindent}        
\def\deg{\ifmmode^\circ \else$^\circ $\fi}    
\def\al{{\it et\thinspace al.}\ }    
\def\ie{{\it i.e.}\ }    
\def\eg{{\it e.g.\ }}    
\def\cf{{\it c.f.\ }}    
\def\kms{km\thinspace s$^{-1}$ }     
\def\amm{\AA\thinspace mm$^{-1}$ }     
\def\arcs{\ifmmode {'' }\else $'' $\fi}     
\def\arcm{\ifmmode {' }\else $' $\fi}     
\def\buildrel#1\over#2{\mathrel{\mathop{\null#2}\limits^{#1}}}
\def\mper{\ifmmode \buildrel m\over . \else $\buildrel m\over .$\fi}
\def\hper{\ifmmode \rlap.^{h}\else $\rlap{.}^h$\fi}
\def\sper{\ifmmode \rlap.^{s}\else $\rlap{.}^s$\fi}
\def\arcsper{\ifmmode \rlap.{' }\else $\rlap{.}' $\fi}
\def\arcmper{\ifmmode \rlap.{'' }\else $\rlap{.}'' $\fi}
\def\gapprox{$_ >\atop{^\sim}$}     
\def\mincir{$_ <\atop{^\sim}$}     
\def\tworule{\noalign{\medskip\hrule\smallskip\hrule\medskip}}
\def\onerule{\noalign{\medskip\hrule\medskip}}
\font\aut=cmbx12
\def\mincir{\ \raise -2.truept\hbox{\rlap{\hbox{$\sim$}}\raise5.truept	
\hbox{$<$}\ }}								%
\def\magcir{\ \raise -2.truept\hbox{\rlap{\hbox{$\sim$}}\raise5.truept	%
\hbox{$>$}\ }}								%
\baselineskip=22pt 

\thispagestyle{empty}
\begin{center}
\null
\bigskip
\bigskip

\bigskip
\bigskip

\Large

{\bf  Fast Spherical Harmonic Analysis:\\ 
a quick algorithm for generating and/or inverting\\
full sky, high resolution CMB Anisotropy maps}

\bigskip
\bigskip
\bigskip

\null
\large

{\it Pio Francesco Muciaccia}\\
{\rm Dipartimento di Fisica, Universit\`a di Roma,``Tor Vergata''}\\
{\rm Via della Ricerca Scientifica, 00133 Roma, Italy}\\
\bigskip
{\it Paolo Natoli}\\
{\rm Dipartimento di Fisica, Universit\`a di Roma, ``Tor Vergata''}\\
{\rm Via della Ricerca Scientifica, 00133 Roma, Italy}\\
\bigskip
{\it Nicola Vittorio}\\
{\rm Dipartimento di Fisica, Universit\`a di Roma, ``Tor Vergata''}\\
{\rm Via della Ricerca Scientifica, 00133 Roma, Italy}\\
\bigskip\bigskip\bigskip
{\it Abstract}\\
\end{center}
\medskip
We present a fast algorithm for generating full sky, high resolution ($\sim5'$)
simulations of the CMB anisotropy pattern.  We also discuss the inverse problem,
that of evaluating from such a  map the full set of $a_{\ell m}$'s and the spectral
coefficients $C_\ell$.  We show that using an Equidistant Cylindrical Projection of
the sky substantially speeds up the calculations. Thus,  generating and/or
inverting a full sky, high resolution  map can be easily achieved with
present day computer technology.
\bigskip\bigskip\bigskip
\begin{center}Subject Headings: cosmic microwave background\end{center} 

\newpage

\section{Introduction}

The angular power spectrum of CMB anisotropies is a gold-mine of  cosmological
information. It sensitively depends upon a number of  parameters: the total and
baryonic density parameters,
$\Omega_0$ and $\Omega_b$; the cosmological  constant, $\Lambda$;  the Hubble
constant,
$H_0$; the spectral indices, $n_s$ and $n_t$ , and the amplitudes  of    scalar and
tensor metric fluctuations; the redshift, $z_{rh}$ , at which the universe could  have
been reionized.  Because of their planned high sensitivity and  high  angular 
resolution, future space missions
will measure the  anisotropy power spectrum with great accuracy. Thus, all these
cosmological  parameters will  be determined with an unprecedented precision
(Bersanelli \al 1996; Jungman \al 1996)

To achieve these goals, Monte Carlo simulations of the CMB anisotropy pattern
have been and will be more  and more crucial in this game. From one hand, they
allow to prepare a mission, to optimize the  observational strategy and to test for
different payload configurations.  On the other hand, they are important  for  the data
analysis and, for example,  to look for systematics.

Up to now, these simulations were realized without problems. Experiments  with
high angular resolution observed only very limited region of the sky: FFT 
techniques   easily provided several realizations of small (and,  hence,  flat) patches
of the sky (see \eg  Kogut, Hinshaw and Bennett, 1995). 
Experiments with large sky coverage, such as COBE/DMR, had low  resolution:
several full sky, CMB anisotropy maps were easily  generated through a  spherical
harmonic expansion with a low ($\mincir 100$) number of  harmonics.

A potential problem it is claimed  to arise in generating even a single high 
resolution,    full sky  map (see, \eg Saez, Holtmann and Smoot, 1996), too much, it is 
believed, for present computer
technology. It  is  generally perceived as a heavy,  almost impossible, computational
task also the inverse problem, that  is extracting out of an observed  full sky, high
resolution map the anisotropy spectrum up to
$\ell \magcir 1000$.  In fact, to our knowledge,   anisotropy  spectrum estimates have 
always been done applying FFT techniques   either  to  small patches of , or even
to   the whole (FFT simulated) sky. The latter approach is  in principle
wrong and its  accuracy   has still to be checked for.

The purpose of this paper is quite technical, but  of interest, 
we believe,  to the  large
community involved in future CMB anisotropy experiments. We want  to present a fast
algorithm for: i) generating   high resolution ($\mincir 10'$), full sky maps; ii)
reconstructing all the coefficients   of a spherical harmonic expansion,
and hence the spectral coefficient $C_\ell$,  up to $\ell
\magcir 1000$. Both these tasks can be easily achieved   on currently available
workstations.  Thus, the plane of this Letter is as follows. In  Sect.2 we will describe
the method. In Sect.3 we will discuss numerical results  and the efficiency of the
method. Finally in Sect.4 we will present a brief  summary of our main findings.

\section{Method}

Generating a CMB anisotropy map is in principle very simple. The temperature
fluctuation observed along a line of sight, $\hat \gamma$,   can be conveniently
described by a spherical harmonic expansions: 
\begin{equation}\label{sum}
{ \Delta T \over T}(\vec x, \hat \gamma) = \sum_{\ell = 0}^{ \ell_{max}}
\sum_{  m= -\ell}^{
\ell} a_{\ell m}(\vec x)  Y_{\ell m}(\hat \gamma)
\end{equation}
where $a_{\ell, -m}= (-1)^m a_{\ell m}^*$. The  
$a_{\ell m}$'s  are random variables of the observer position, $\vec x$, gaussian
distributed (at least in most of the inflation based scenarios), with zero mean and
variances
$\langle|a_{\ell m}|^2\rangle \equiv C_\ell$.  In simulating the CMB, primary
anisotropy pattern the sum over $\ell$ usually starts from two. In fact,
from one hand  the monopole   vanishes
  by construction, being the mean (over the sky) CMB anisotropy. On the other
hand,   the dipole components  are  dominated by the Doppler anisotropy,  induced
by our peculiar motion relative to the comoving frame (see \eg Kogut \al 1994).  
On the same line, we will keep the sum over  $\ell$ from $ 0$ to
$\ell_{max}$, but we set     
$a_{00}$  =
$a_{1,-1}$ =
$a_{1,0}$ =
$a_{1,1}$ = $0$.  

The
$C_\ell$'s are the main prediction  of a theory of structure formation (see \eg Hu
and Sugiyama, 1996). Thus, for a given scenario (\ie for given $C_\ell$'s)  and for a
given statistics,
 we can generate a random set of $a_{\ell m}$'s and, from Eq.(\ref{sum}), a CMB anisotropy
map. In practice, using the spherical harmonic expansion as in Eq.(\ref{sum}) is not very
efficient: for each line of sight
$\hat\gamma$ we should  evaluate $Y_{\ell m}(\hat\gamma)$ for each value of
$\ell$ and $m$. Fortunately, it is possible to rewrite  Eq.(\ref{sum}) in a form more suitable for
numerical implementation.

First, it  is easy to verify that the double sum in Eq.(\ref{sum}), 
 $ \sum_{l=l_0}^{l=l_{max}}
\sum_{m=-\ell}^{m=+\ell} $,  is completely equivalent to  $
\sum_{m=-\ell_{max}}^{m=+\ell_{max}}
\sum_{\ell=|m|}^{\ell= \ell_{max}}$:   we sample exactly the same region of the
$\ell-m$ space, by columns  (in the former) or  by rows (in the latter   case; \cf
Fig.1).   Second, let us  write
\begin{equation}\label{Y}
Y_{\ell m}(\theta,\phi)=\lambda_\ell^m(\cos\theta) e^{im\phi},
\end{equation}  
where
\begin{equation}\label{lambda}
\lambda_\ell^m=\sqrt{{2\ell +1 \over 4\pi} {(\ell-m)!\over (\ell+m)!}} 
P_\ell^m(\cos\theta),
\end{equation}
 $\lambda_\ell^{- m } = (-)^m \lambda_\ell^{m}$   and
$P_\ell^m(\cos\theta)$ are the associated Legendre polynomials.  It is then possible
to rewrite Eq.(\ref{sum}) as follows:
\begin{equation}\label{sum2}
 { \Delta T \over T}(\phi,\theta) = \sum_{m=-\ell_{max}}^{m=+\ell_{max}}
b_m(\theta) e^{im\phi}
\end{equation}
 where 
\begin{equation}\label{b}
b_{ m }=\sum_{\ell=|m|}^{\ell= \ell_{max}} a_{\ell  m }
\lambda_\ell^{ m }
\end{equation}
and $b_{- m }=b^*_{ m }$. Written in this way, Eq.(\ref{sum2}) highlights a couple of
attractive features.  If we  use an Equidistant Cylindrical Projection (hereafter
ECP) of the sky,  which conserves distances along meridians and along the equator, 
the anisotropy  map can be 
thought as a rectangular matrix of   $N_\phi$  times
$N_\theta(=N_\phi/2)$ squared pixels, each  of  dimension $\simeq 20'
\times 20' (1024/N_\phi)^2$.  In this projection, the temperature
anisotropy along parallels (\ie at fixed
$\theta$) is nothing more than the 1D Fourier transform of the coefficients
$b_m(\theta)$'s [\cf Eq.(\ref{sum2})],   very efficiently computed with FFT
techniques.   If we regard the sum of Eq.(\ref{sum2}) as a Fourier expansion, then
$\ell_{max}$  must be  fixed to be $N_\phi/2$, as it  plays the role of the
Nyquist critical frequency of the problem. Second, the $\lambda_\ell^m$ are evaluated
by standard 
recurrence relations:
\begin{eqnarray}
\lambda_m^m &=& (-1)^m  \sqrt{{2m+1\over 4\pi}}{(2m-1)!!\over \sqrt{(2m)!}}
(1-x^2)^{m/2}\nonumber\\
\lambda_{m+1}^m &=& x \sqrt{2m+3}\lambda_m^m\nonumber\\
\lambda_\ell^m &=& \biggl[x\lambda_{\ell-1}^m -
\sqrt{{(l+m-1)(l-m-1)\over
(2\ell-3)(2\ell-1)}}\lambda_{\ell-2}^m\biggr]\sqrt{{4\ell^2-1
\over
\ell^2-m^2}}\label{rec}
\end{eqnarray} 
where $x=\cos\theta$. Because of these relations,    the
$b_m(\theta)$'s can be computed  very efficiently, as we can  perform 
the sum
in Eq.(\ref{b}) while    computing the  
$\lambda_{\ell }^m$'s. Such a computation is further simplified because
$\lambda_{\ell m}(\cos\theta)=\pm \lambda_{\ell m}[\cos (\pi-\theta)]$, 
the plus  (minus) sign holding  if  $\ell$ and $m$ are  (are not) both even or
 both odd. 

Generating the  $b_m(\theta)$'s at fixed $\theta$  requires evaluating $\approx
\ell_{max}^2 $ $(\propto N_\phi^2)$   recurrence relations  ($\ell_{max} - m$
recurrence relations  for each value of $m$). The CPU time needed for FFT-ing the
$b_m(\theta)$'s in principle scales as $N_\phi\ln N_\phi$. However, the FFT is so
fast that for $N_\phi \leq 4096$ most of the time is spent for generating the
$b_m(\theta)$'s. Finally, we have to evaluate the $b_m(\theta)$'s
$N_\theta$ times. So, at the end, the total CPU time needed for
generating an ECP of the anisotropy pattern is expected to  scale  as
$N_\phi^3$. 

At this point  it is quite easy to address  also the inverse problem, that of evaluating
from an observed high resolution, full sky map   the set of coefficients $a_{\ell m}$.
It is very well known that the orthonormality of the spherical harmonics allows to
invert Eq.(\ref{sum}) and write:
\begin{equation}\label{alm}
 a_{\ell  m }= \int d\Omega  {\Delta T\over T}(\hat \gamma) Y^*_{\ell m}(\hat
\gamma)
\end{equation}
This sounds awful: in principle for each $\ell$ and $m$ we should evaluate
$Y_{\ell m}(\hat \gamma)$ for a given $\hat\gamma$  and integrate over the whole
sky.  Fortunately, after substituting Eq.(\ref{Y}) in Eq.(\ref{alm}) we can write:
\begin{equation}\label{alm2}
a_{\ell m}= \int \sin\theta d\theta \lambda_\ell^m (\theta) b_{m}(\theta)
\end{equation}
where 
\begin{equation}\label{b2} 
b_{m}(\theta)=\int_0^{2\pi} d\phi \Delta(\phi,\theta) \exp(-i m \phi)
\end{equation}
Thus,  Eq.(\ref{alm2}) is the conjugate of
Eq.(\ref{sum2}): the $b_m$'s are the Fourier anti-transform of the anisotropy pattern 
along a parallel in the ECP of the sky, and are easily
computed at fixed $\theta$ with a 
FFT.   In conclusion, inverting a map to obtain  the  $a_{\ell m}$'s
requires
$\approx
\ell_{max}^2$ ($\propto N_\phi^2$) recurrence relations for evaluating the
$\lambda_\ell^m$'s, plus a FFT to evaluate the $b_{ m}$'s. All this must be done
$ N_\theta$ ($\propto N_\phi$) times to be able to perform the integral in Eq.(\ref{alm2}).
  Using these tricks, we can invert a full sky, high resolution map
with CPU times which scale as
$N_\phi^3$ (the evaluation of the
$b_{\ell m}$'s is basically instantaneous), and in principle comparable with those
needed for generating a map. As in that case, the actual CPU time can be further
reduced by exploiting the simmetries of the $\lambda_{\ell m}$'s evaluated at
$\theta$ and $\pi-\theta$, respectively.

\section{Numerical Results}

In the previous Section we described an algorithm for generating and/or inverting a
  high resolution, full sky map of the CMB anisotropy.  In this Section we will briefly
discuss the actual
 performances of this algorithm on a DEC 1000/200. We will consider  hereafter 
the  standard Cold Dark Matter model.

In Fig.2 we plot the CPU time needed for generating  a  full sky,
ECP of the CMB anisotropy pattern as a function of  the angular  resolution, the
only free parameter we can play with. In fact, we sample the ECP of the sky with
$N_\phi\times N_\theta$ squared pixels of dimension $20'\times 20'
(1024/N_\phi)^2$. This fixes $N_\theta=\ell_{max}=N_\phi/2$.  The expected
scaling with $N_\phi^3$ (see Sect.2) is recovered with good precision. 
We want to stress that  only 1h of CPU time  is needed to
generate a full sky, ECP  of the CMB anisotropy  with a resolution of $5'$
(\ie
$N_\phi=4096$).  

Using an ECP is not a limitation. In fact, once  we obtain an
ECP of the anisotropy pattern, we can reproduce it in any given projection. As an
example we show in  Plate 1 an ECP, 10' resolution map (obtained in only $8$
minutes of  CPU time) and the corresponding Equal Area Projection (hereafter EAP). 
The latter is obtained by the former using standard spherical 
trigonometry. We verified that this  procedure is numerically stable. 
In fact, if we transform an ECP to an EAP  and the obtained EAP  back to an ECP, we
reproduce the initial  anisotropy pattern exactly.  Thus, from
observations of the CMB anisotropy we can create an ECP of the sky
  and then  apply our inversion algorithm.  In Fig.2 we also show the CPU time
needed for inverting an ECP  map as a function of the map resolution. 
The CPU time scales
roughly as
$N_\phi^3$. Again, $\simeq 1h$ of CPU time is needed to recover from 
a $5'$ resolution map the entire set of $a_{\ell m}$'s, roughly 
$4\cdot 10^6$ coefficients.

In Figs.3 and 4 we show the precision of our inversion algorithm for
$\ell_{max}=1024$ (corresponding to a resolution of $10'$) .  The percentage error
between the recovered
$a_{\ell m}$'s and the input ones  is large  only  for
$\ell \sim   \ell_{max}$ and $m\sim  $ few, a very small portion of the
allowed region of the $\ell-m$ space. This is due to the fact that for
$m
\rightarrow 0$,   
$\lambda_\ell^0  \propto P_\ell (\cos\theta)$: for large values of $\ell$,  this is
a  highly oscillating function of
$\theta$. On the contrary,   for
$m\rightarrow
\ell$, 
$\lambda_\ell^\ell  \propto 
  (1-\cos\theta)^{\ell/2 }$, a quite smooth function of the azimuthal angle [\cf
Eq.(\ref{rec})]. So, a simple trapezoidal rule for performing the integral along meridians in
Eq.(\ref{alm2}) gives a poor result  only for very  large values of $\ell$ and quite small values
of
$m$.  However, this is not a crucial problem. In fact, we are mostly interested in
evaluating the spectral coefficients
$C_\ell$. These are obtained from the recovered $a_{\ell m}$'s as follows:
\begin{equation}\label{c_l} 
C^{estimated}_{\ell} = {1\over  2\ell +1}  \sum_{m=-\ell}^{\ell} |a_{\ell m}|^2
\end{equation} 
It is clear that the error we make in recovering the $a_{\ell m}$'s for large $\ell$ and
small $m$ is highly  diluted in  the sum of Eq.(\ref{c_l}).  In Fig.5  we show the
(percentage) error between the recovered and the input $C_\ell$'s as a function,
again,  of the map resolution. The recovered spectrum has a maximum error of
$\sim 0.1\%$   up to $\ell\mincir 1500$ for
$N_\phi =4096$, \ie for a pixel size of $5'\times 5'$. 
 
\section{Conclusions}

We present a fast algorithm  for: i) generating high resolution, full sky maps of the
CMB anisotropy; ii)    evaluating out of an observed map 
the $a_{\ell m}$'s and then the spectral coefficient $C_\ell$'s.  The basic trick for
speeding up the calculation consists in generating and/or inverting a full sky map
using an ECP. It is this 
projection that allows the use of a FFT either in Eq.(\ref{sum2})
and/or in Eq.(\ref{b2}).  If we are interested in probing the anisotropy power spectrum up
to 
$\ell_{max}
\simeq 1000$, then $N_\phi= 2\ell_{max} =2048$ and we need only $8$ minutes of 
CPU time for either generating or inverting a $10'$ resolution, full sky map.
Pushing the sampling down to $5'$ boosts the  needed CPU time up
to  one hour. Our algorithm also allows us to fully exploit a parallel
architecture,  such as the one of APEmille (Bartoloni \al 1995). This machine is
composed  by 1024 processors, each of them slower by roughly a factor of two w.r.t. a
DEC 1000/200. So, with  such a machine  one should be able to produce 1024, $5'$
resolution maps of the anisotropy  pattern in a couple of hours. Details about
this application will be discussed elsewhere.
 
In addressing the problem of inverting a full sky map, we assumed full 
sky coverage and we fully exploited the orthonormality of the spherical 
harmonics. We test our algorithm against a pure CMB anisotropy pattern. 
In the realistic case, a $\mu$-wave map will be the superposition of 
different processes (CMB anisotropy, Galactic foregrounds, secondary 
anisotropy due to clusters of galaxies, point sources, etc.) and the sky coverage can 
be not complete. The  separation of the CMB anisotropy pattern from Galactic and 
extragalactic foregrounds has been studied in details (Bouchet \al 1994; Bouchet
\al 1995; Bersanelli \al 1996), but considering only small ($10^o\times  10^o$)
patches of the sky. We will discuss an application of our  algorithm to the problem of
foreground subtraction and not complete sky coverage in a forthcoming  paper.

\newpage
\begin{figure}
\begin{center}\mbox{\epsfxsize=8cm\epsffile{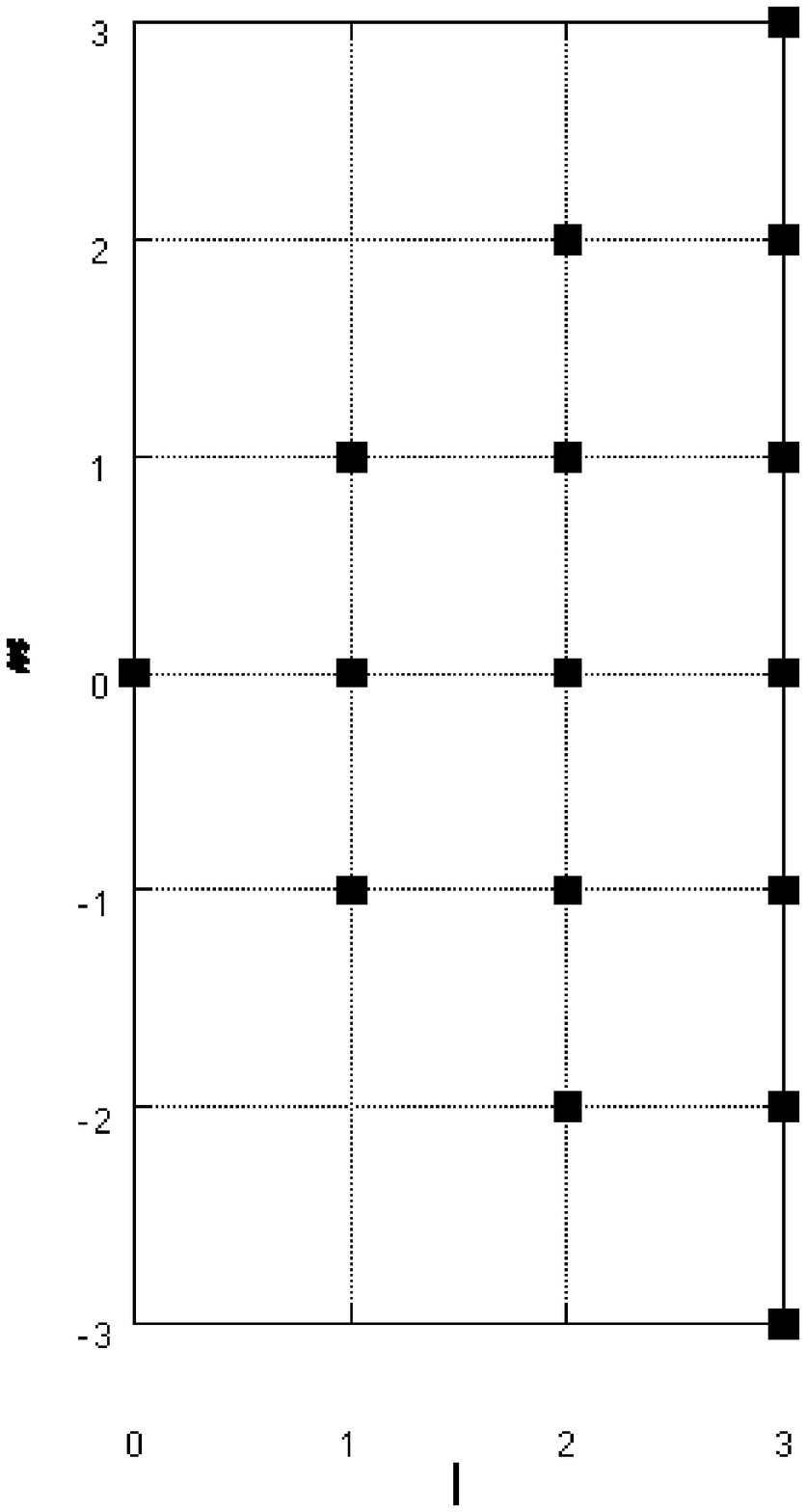}}\end{center}
\caption{The filled squares show  (for $\ell_{max}=3$) the portion of the
$\ell-m$ space probed by the $a_{\ell m}$'s. We can recover all the allowed 
$\ell- m$
pairs either by moving along columns ($\sum_{\ell = 0}^{ \ell_{max}}
\sum_{  m= -\ell}^{\ell}$) or by rows ($\sum_{m=-\ell_{max}}^{m=+\ell_{max}}
\sum_{\ell=|m|}^{\ell= \ell_{max}}$).}
\end{figure} 

\begin{figure}
\begin{center}\mbox{\epsfxsize=15cm\epsffile{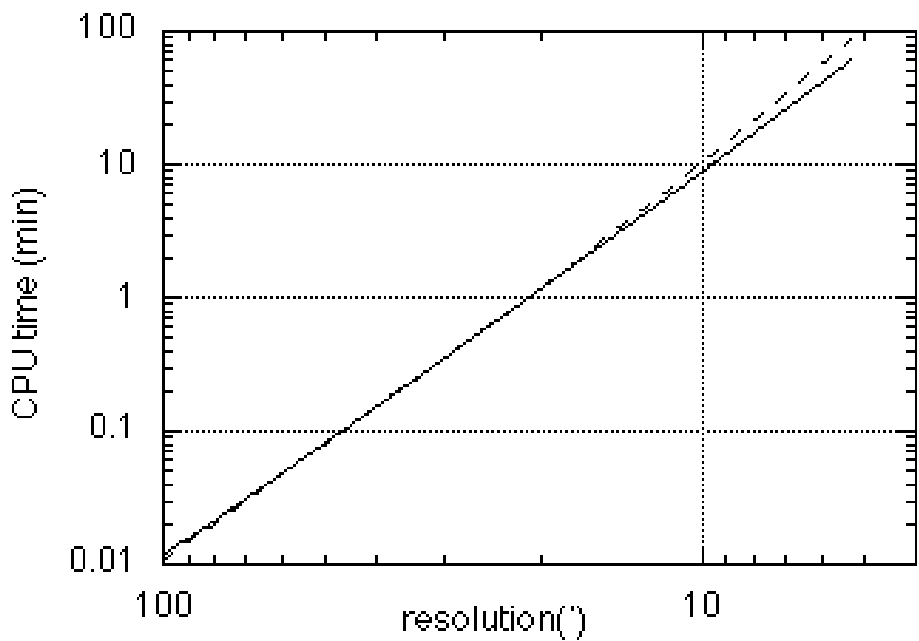}}\end{center}
\caption{The CPU time needed, on a DEC 1000, to generate a full sky map (solid
line) or to invert a map to generate the full set of $a_{\ell m}$'s (dotted
line) as a function of the sampling of an Equidistant Cylindrical Projection 
of the sky.}
\end{figure} 

\begin{figure}
\begin{center}\mbox{\epsfxsize=10cm\epsffile{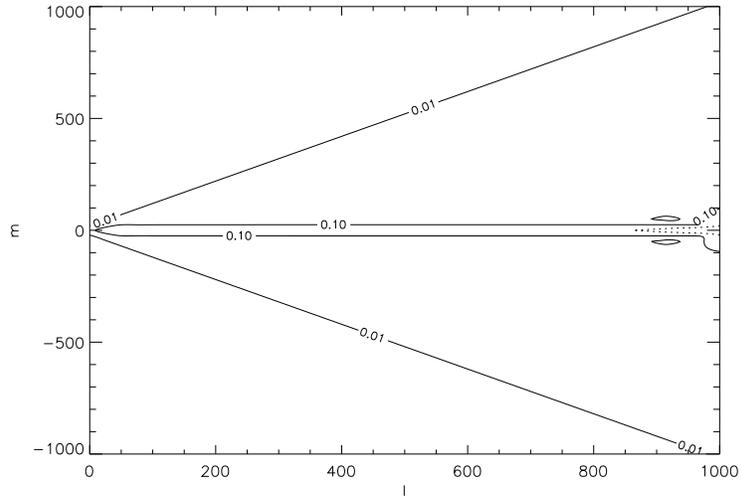}}\end{center}
\caption{The percentage error between  the recovered   
 and the input, real part of
the $a_{\ell m}$'s.  The labels indicate the $0.01\%$ and the $0.1\%$ isolevel,
respectively. Only in  a very small portion of the
$\ell-m$ plane  the error is larger than 10\% (dashed line).}
\end{figure} 

\begin{figure}
\begin{center}\mbox{\epsfxsize=10cm\epsffile{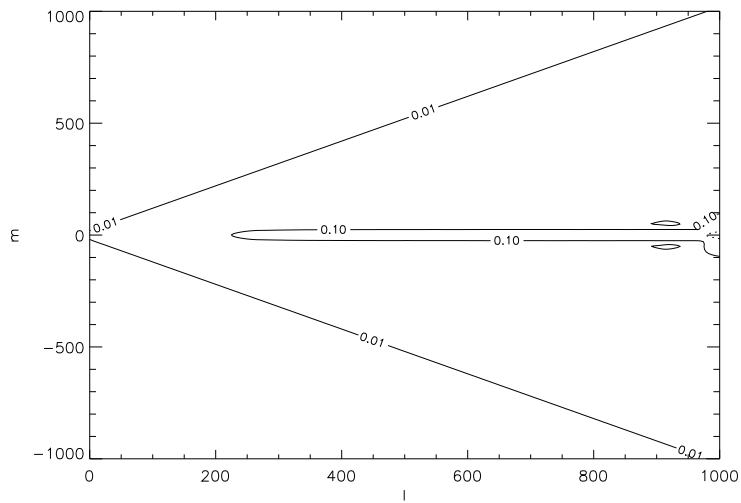}}\end{center}
\caption{Same as in Fig.3, but for the imaginary part of the $a_{\ell 
m}$'s.}
\end{figure} 

\begin{figure}
\mbox{\epsfxsize=14cm\epsffile{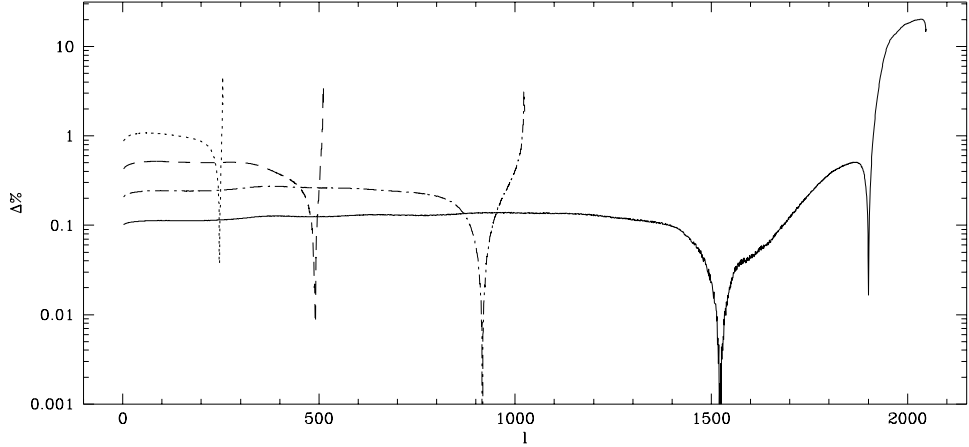}}
\caption{The percentage error between the recovered and the input  spectral 
coefficients 
$C_{\ell}$'s as a function of $\ell$ for different resolutions of the 
Equidistant Cylindrical Projection of the CMB anisotropy pattern. Dotted, dashed,
dot-dashed and continous lines refer to map resolution of 
$\simeq 40'$ ($\ell_{max}=256$),
$\simeq 20'$ ($\ell_{max}=512$),
$\simeq 10'$ ($\ell_{max}=1024$),
$\simeq 5'$ ($\ell_{max}=2048$), respectively.}
\end{figure} 

\begin{figure}
\caption {A realization of the CMB anisotropy pattern in a standard 
Cold Dark Matter model in the 
Equidistant Cylindrical (panel a) 
and in the Equal Area (panel b) Projections.}
\end{figure} 

\end{document}